\documentclass[12pt,a4paper,oneside]{article}
\usepackage{epsfig}
\usepackage{pstcol}
\def\PL #1 #2 #3 {{\it Phys.~Lett.~} {\bf#1} (#3) #2}
\def\NP #1 #2 #3 {{\it Nucl.~Phys.~} {\bf#1} (#3) #2}
\def\ZP #1 #2 #3 {{\it Z.~Phys.~} {\bf#1} (#3) #2}
\def\PRL #1 #2 #3 {{\it Phys.~Rev.~Lett.~} {\bf #1} (#3) #2}
\def\PR #1 #2 #3 {{\it Phys.~Rev.~} {\bf#1} (#3) #2}
\def\MPL #1 #2 #3 {{\it Mod.~Phys.~Lett.~} {\bf#1} (#3) #2}
\def\RMP #1 #2 #3 {{\it Rev.~Mod.~Phys.~} {\bf#1} (#3) #2}
\def\EPJ #1 #2 #3 {{\it Eur.~Phys.~J.~} {\bf#1} (#3) #2}

\def\ifm{\ifmmode}

\definecolor{Pink}{rgb}{1.,0.75,0.8}
\newcommand{\beq}{\begin{equation}}
\newcommand{\eeq}{\end{equation}}
\newcommand{\beqn}{\begin{eqnarray}}
\newcommand{\eeqn}{\end{eqnarray}}
\newcommand{\beqs}{\begin{eqnarray*}}
\newcommand{\eeqs}{\end{eqnarray*}}

\newcommand{\pdf}{PDF\ }
\newcommand{\pdfb}{PDF}
\newcommand{\pdfs}{PDF's\ }
\newcommand{\pdfsb}{PDF's}
\newcommand{\W}{$W$-boson\ }                   
\newcommand{\Z}{$Z$-boson\ }                   
\newcommand{\F}{\mbox{$\cal F$}}                   
\newcommand{\D}{\mbox{$\cal D$}}                   
\newcommand{\alfa}{\mbox{$\alpha_S$}}
\newcommand{\QCD}{QCD\ }
\begin{document}
\thispagestyle{empty}
\vskip 1cm
\begin{center}
{\Large Parton Distribution Function Uncertainties}
\vskip 1cm
Walter T.~Giele\\
{Fermi National Accelerator Laboratory, Batavia, IL 60510} \\
\vskip .6cm
St\'ephane A.~Keller\\
{Theory Division, CERN, CH 1211 Geneva 23, Switzerland} 
\footnote{Supported by the European Commission under
    contract number ERB4001GT975210, TMR - Marie Curie Fellowship}\\ 
\vskip .6cm
David A.~Kosower\\
{CEA--Saclay, F--91191 Gif-sur-Yvette cedex,
    France.}\\
\end{center}
\vskip .3cm
\begin{abstract}
We present parton distribution functions which include a quantitative estimate 
of its uncertainties. The parton distribution 
functions are optimized with respect to deep 
inelastic proton data, expressing 
the uncertainties as a density measure over the functional space of
parton distribution functions. This leads to a convenient method of 
propagating the parton distribution function uncertainties to
new observables, now expressing the uncertainty
as a density in the prediction of the observable. 
New measurements can easily be included in the optimized sets
as added weight functions to the density measure.
Using the optimized method nowhere in the analysis 
compromises have to be made with regard to
the treatment of the uncertainties.
\end{abstract}

\newpage
\section{Introduction}

With the advent of new hadron collider experiments the need to quantitatively
estimate parton distribution function (\pdfb) uncertainties is of paramount importance. One
obvious example where the \pdf uncertainty will become crucial is the \W mass uncertainty.
By no means is this the only observable for which the \pdf uncertainty is 
of the utmost importance. 

The realization that hadron collider experiments had reached
a level of accuracy where a quantitative approach
towards the \pdf uncertainties was needed came 
with the one-jet inclusive transverse energy distribution measurement of the CDF
collaboration using the run 1a data with an integrated luminosity of 
19.5 pb$^{-1}$~\cite{onejet1a}. Up to that point the qualitative ``global fitting''
approach had worked well and was used to demonstrate the success of the perturbative
\QCD framework. Be that as it may, the precision of the CDF one-jet inclusive measurement 
demonstrated that the qualitative approach had reached its limit in usefulness.
The fact that prior to the measurement none of the \pdfs could predict the high 
transverse momentum data whereas after the measurement the \pdfs could be adjusted
to accommodate the high transverse energy excess~\cite{onejetCTEQ} speaks for itself.
The only way to continue is to use a quantitative approach which reflects 
the \pdf uncertainties.

The qualitative ``global fitting'' approach~\cite{mrs99,cteq5m} combines a large amount of experimental
data to form an error weighted average of all input data, i.e. find the 
minimum $\chi^2$ solution. In the calculation of the $\chi^2$ the experimental statistical 
and systematic uncertainties are added in quadrature. 
Because experiments are combined with each 
other, even if no statistical adequate solution could be found,
no conventional probabilistic interpretation can be reached (see for instance section II.D of 
ref.~\cite{PDFsnowmass} for an overview of the paradoxal problems encountered
when attempting a probabilistic interpretation in the ``global fitting'' approach).

It is crucial to develop a rigorous statistical approach such that the uncertainties have 
an objective statistical interpretation and in case of deviations the experiments 
or prior assumptions causing the conflict are traceable.
The apparent way to proceed
would be to redo the ``global fitting'' approach with the uncertainty
analysis in mind. That is, treating the experimental error analysis more carefully
by including correlation matrices reflecting the experimental uncertainty
and by demanding the input experiments to be compatible.
As an advantage this would still allow a $\chi^2$ calculation with a minimalization
procedure. The final result is then quoted as a correlation matrix of the \pdf parameters,
i.e. a 2nd order Taylor expansion around the best solution. Yet, this method has problems.
First of all the experimental systematic uncertainties are not gaussian despite the
fact that often gaussian approximations are published. Secondly, one can
expect complicated correlations between the \pdf parameters given the fact that the problem
of \pdf fitting is highly non-linear causing lower dimensional non-trivial regions 
in \pdf parameter space with constant probability measure (i.e. constant $\chi^2$).
Thirdly, in the end the \pdfs are not the physical observables and propagation from
the \pdf parameters to the physical observable would again necessitate a linearization
of the error propagation involving
an estimate of the derivative of the observable with respect to the
\pdf parameters. In short, this method has three layers of subsequent linearizations
of uncertainties which can lead to problems.

The first paper using this method ~\cite{alekhin} uses the
H1, ZEUS, BCDMS and NMC data. Subsequently, refs.~\cite{botje,zomer} 
include additional experiments into the fits.
Recently, a Lagrange multiplier method was outlined which does not need to make
gaussian approximations~\cite{CTEQlagrange}.
Also proposed in refs ~\cite{CTEQlagrange,CTEQhessian}
is a procedure to maintain the ``global fitting'' philosophy by
include as many experiments as possible. 
To recover a probabilistic interpretation one has to 
multiply the experimental uncertainties with a common factor such
that all experiments are consistent within a single \pdf ``global fit''.

Parallel to the previous methods, a different approach was pursued 
in ref.~\cite{gielekeller}
to remove the last of the three gaussian approximations in the error analysis.
Using the gaussian parameter probability density of ref.~\cite{alekhin} 
an optimized Monte Carlo integration approach 
was applied to predict physical observables
expressing its \pdf uncertainty as a density probability measure in the predictions
of the observable. Note that such an approach can be mathematically reformulated as a
statistical inference method.
In this paper we extend this method to eliminate all approximations
in the error analysis. As a result any experimental error analysis can be implemented
and the final resulting uncertainties for observables can assume any form. The intermediate
\pdf parameter probability distribution can have any shape required by the experiments.
The resulting optimized Monte Carlo approach is transparent, easy to use, modular and
extendible. These properties are important for practical use.

In section 2 we will explain our method of optimized Monte Carlo integration.
After that we are ready to optimize the \pdfs towards sets of experiments.
In section 3 we will consider the proton $F_2$ results of
H1, ZEUS, BCDMS and NMC together with an world averaged value of $\alpha_S(M_Z)$.
The conclusions and outlook will be given in section 4.
\section{The Method}

This section will describe the methodology we developed to obtain theoretical
predictions of physical observables including \pdf uncertainties. 
Before we can formulate our method we need to define carefully
all aspects of comparing experiments and theory.  
First of all, one has to define a theoretical framework which approximates
the true nature value of observables.
Secondly, the experimental results have to be casted in a well defined object which 
will take the form of an experimental response function.
After defining the theoretical prior and the experimental response function 
we have the right language to formulate our method.

\subsection{The Model Priors}

Any observable has a {\it true nature} value $x_T$ which has to be calculated
(i.e. approximated)
by a theory model giving a value $x_t$. The \pdfs only consist within the framework of
the parton model and perturbative \QCD. This implies approximations have been made to
arrive at this calculable model. Hence the resulting \pdfs will depend on the theory
model defined in the prior. As long as the experimental uncertainties are larger
than the deviations from the {\it true nature} value due to the theory model 
the approach is valid. The model uncertainties fall into three distinct classes. 

The first class is an agglomeration of non-perturbative effects. 
One has to make the approximation in which factorizable
\pdfs and calculable hard scattering matrix elements exist~\cite{CSS}. This leads to
the appearance of power suppressed momentum transfer terms. These higher twist terms
can be either parametrized (see e.g. ref.~\cite{botje}) or assumed small enough 
to be neglectable by applying appropriate kinematic cuts. Moreover,
the final state outgoing partons are assumed to have factorizable fragmentation functions.
This again leads to power suppressed momentum transfer terms. Finally there is
the interaction of the colliding hadron remnants with the hard scattering. For
this effect no real physics models exist and usually the experiment is ``corrected''
for the ``underlying event''  fathering a large systematic uncertainty. 
For this paper we will neglect all such effects in the theory prior
and assume they are neglectable with
respect to the other uncertainties in play for the kinematic cuts applied.
The underlying event subtraction is, as usual, considered an experimental problem.
One final non-perturbative effect comes in play when experiments are considered which
use other initial state hadrons than protons (e.g. deuterium, iron, copper, etc.). In such
a case one has to consider the effect of the multitude of nuclei. That is, what is the
relation between the free proton \pdf and the measured atomic densities. In this paper
we will not include heavy target data and therefore can ignore the phenomelogical shadow
models which attempt to describe these nuclear effects.
Note that the inclusion of non-perturbative models will involve the introduction 
of additional degrees of freedom (i.e. non-perturbative parameters). The effect is to
de-emphasize certain kinematical regions associated with low momentum transfer scattering. 
Such non-perturbative models, if included in the theory prior, 
must be applied consistent to all predictions.

The second class of uncertainties is related to the
perturbative expansion of the hard scattering partonic cross sections. 
Because this is in principal calculable the uncertainties are more traceable.
In the theory prior one has to determine the order in perturbative \QCD to which
all observables will be calculated. This also defines the order to which the \pdf
evolution has to be performed. A special case would be to define a resummation scheme
of dominant initial state logarithms to extend 
the kinematic range of reliable predictions. If such a
scheme is adopted it should be applied to {\it all} predictions. Also such a scheme could
affect and modify the evolution equations. For this paper we choose the 
next-to-leading order approximations in the $\overline{\mbox{MS}}$ 
renormalization/factorization scheme for
all observables. For the appropriate evolution of the \pdfs the program
{\tt QCDNUM} was used~\cite{qcdnum} which accuracy is more than adequate for current
phenomenology~\cite{Evolution}. Note a theoretical uncertainty could be
assigned to the fixed order hard scattering matrix element calculation 
which reflects an estimate of the deviation from the prediction to an even 
higher order calculation of the observable. 
This can take many forms and will be highly subjective. However, one can define such
a prior and incorporate it easily in the uncertainty analysis.
In this paper we restrict ourselves to only looking at the 
renormalization/factorization scale dependence of the next-to-leading order
theory predictions. To do this we allow a floating renormalization/factorization
scale in the optimalization procedure. Given enough orders in the perturbative expansion
virtually no scale dependence would remain. However, at next-to-leading order there will
be a scale dependence on the observables and therefore the \pdfsb. 
The extend to which the scale can vary independent of
the \pdf probability measure indicates the need to increase the perturbative order of
the matrix element calculation. Given enough experimental accuracy the fixed order
calculation will fail. This will be reflected through the scale in a strong
preference for a specific value of this scale.
 
The final class of model priors is a set of additional requirements. 
Some of them can be physics
necessities, e.g. the \pdf charge and momentum conservation sum rules. Others can
be more speculative, for instance assumptions about small and large parton fraction 
behavior or introducing \pdf moment constraints from lattice \QCD 
(see e.g.~\cite{lattice}). 
Also, one could use a previous \pdf optimalization as a prior and include
additional experiments in the optimized \pdf sets. Note that this requires that
the prior optimalization results are consistent with the new experiments, i.e. that
the inclusion
leads to a refinement of the relevant region contributing to the functional integration
in the functional space $V(\F)$ of \pdfs $\F$. 
The most important constraint in this class are smoothness constraints on the 
actual \pdfs which must be introduced to regulate unconstrained 
fluctuations due to the discrete nature of experimental results. 
The most restrained introduction of some of these requirement
is to define a specific parameterization for the functional form of the \pdfs 
depending on a fixed, finite number of parameters. Note however that this still
requires an assumption on the initial probability density distribution of those parameters.
For this paper we choose one such parameterization as is detailed in the next section.
All in all,  the constraints have to be quantified and assembled
as a prior probability density function measure, $P_{prior}(\F)$, 
over the \pdf functional space $V(\F)$.

\subsection{The Experimental Response Function}\label{ERP}

Given the theory prior we now can make well defined approximations $x_t$ of the
{\it true nature} value $x_T$ for any relevant observable.
In order to interact with the experimental results we have to define 
the experimental response function
\beq
P_{exp}(x_e|\D(x_t))=P_{exp}(x_e|x_o)\ ,
\eeq
which is a probability density estimating the likelyhood of measuring $x_e$ given the 
theory prediction $x_t$ and detector response ${\cal D}$. The detector response
is the deformation inflicted on the physics signal by the detector
resulting in the measured signal $x_o=\D(x_t)$.
The systematic uncertainties are now defined as a
probability density measure, $P_{sys}(\D)$,
over the functional space $V(\D)$ of all potential detector responses for the observable
and quantifies the understanding of the detector.
Note that in principle the experimental response function has to be formulated
before the actual measurement of the particular observable. Also, theory predictions
can originate from any source and are not tied to a specific model in any way.

Often the detector response is specific to a particular measurement and not
correlated with any other experiment under consideration. In that case we
can integrate over the detector response functional space
\beq\label{sys}
P_{exp}(x_e|x_t)=\int_{V(\D)}d\,\D\ P_{exp}(x_e|\D(x_t))\times P_{sys}(\D)\ ,
\eeq
thereby simplifying the analysis. For instance, if the detector response probability function
is parameterizable in a multi-gaussian the above integration would simplify to the usual
$\chi^2$ formulation with a correlation matrix encapsulating the systematic uncertainties. 

Note that if several measurements with correlated systematic uncertainties are
included one has to integrate the systematic uncertainties over the group of 
correlated measurements.
A more flexible approach is to use the optimized Monte Carlo approach 
(explained in the next subsection) not only for
the \pdfs but in conjunction with the detector response \D. This way, each \pdf has a
specific detector response \D\ which makes propagation of systematic uncertainties to
other observables trivial.

The experimental response function $P_{exp}$ together with the detector response function
$P_{sys}$ contains all information we can extract from the measurement. Together with
the actual measured values of the observable, $x_m$, 
they form a permanent and well defined record of the measurement.

\subsection{The Optimized Monte Carlo Approach}

We now can formulate our method. To make a prediction for observable ${\cal O}$
which includes the \pdf uncertainty based on prior assumptions and certain 
sets of measurements, we perform the integration over the \pdf and detector response 
functional space to obtain the probability density function of observing a value
$x_e$ for observable ${\cal O}$
\beq\label{Ifunctional}
P_{pdf}^{\cal O}(x_e) = \int_{V(\F)} d\,\F\ P_{exp}^{input}(\F)\times P_{prior}(\F)\times 
P_{exp}^{\cal O}\left(x_e|x_t(\F)\right)\ ,
\eeq
where $P_{exp}^{\cal O}\left(x_e|x_t(\F)\right)$ is the experimental response function 
for observable ${\cal O}$ using \pdf set \F. Note we have integrated out the systematic
uncertainties for the observable as explained in eq.~\ref{sys} and thereby assuming its
systematic uncertainties are independent of the other experiments involved in the \pdf
determination. 
The prior probability function $P_{prior}$ is defined in subsection 2.1 and quantifies
all prior \pdf assumptions and regulators. Finally, the probability function 
of the input experiments is defined as
\beq\label{Pfunctional}
P_{exp}^{input}(\F) = \int_{V(\D)}d\,\D\  P_{exp\ 1}(\F,\D)\times P_{exp\ 2}(\F,\D)
\times\cdots\times P_{exp\ n}(\F,\D)\times P_{sys}(\D)\ ,
\eeq
with the individual experimental response function for  each experiment
\beq
P_{exp\ k}(\F,\D)=P_{exp\ k}\left(x_m^{(k)}\Big|\D\left(x_t^{(k)}(\F)\right)\right)\ ,
\eeq
where the actual measurement $x_m^{(k)}$ is substituted 
in the experimental response function.
While formally this defines the \pdf uncertainty for any observable we want,
in practice a numerical integration over the \pdfs and possible systematic uncertainties
has to be implemented. The only feasible method in this situation is a Monte Carlo approach.
That is, we approximate the integral by randomly picking a sufficient large set of $N$
\pdfs and possible detector response functions such that
\beq\label{Isum}
P_{pdf}^{\cal O}(x_e)\approx\frac{1}{N}\sum_{i=1}^N P_{exp}^{input}(\F_i)\times P_{prior}(\F_i)\times 
P_{exp}^{\cal O}\left(x_e|x_t(\F_i)\right)\ ,
\eeq
and
\beq\label{Psum}
P_{exp}^{input}(\F_i) = P_{exp\ 1}(\F_i,\D_i)\times P_{exp\ 2}(\F_i,\D_i)
\times\cdots\times P_{exp\ n}(\F_i,\D_i)\times P_{sys}(\D_i)\ .
\eeq
While this can be numerically implemented, the Monte Carlo integration efficiency defined
as the ratio of the average of the combined probability function and the maximum of the
combined probability function 
\beq
\epsilon(P(\F,\D))\approx\left(\frac{1}{N}\sum_{i=1}^N P(\F_i,\D_i)\right)
\Big/ \left(\max_i P(\F_i,\D_i)\right)\ ,
\eeq
with
\beq
P(\F,\D)=P_{exp\ 1}(\F,\D)\times P_{exp\ 2}(\F,\D)\times\cdots\times P_{exp\ n}(\F,\D)
\times P_{sys}(\D)\times P_{prior}(\F)\ ,
\eeq
is for all practical purposes zero. This because given the virtually infinite functional
space of choices of \F\ and \D\ the likelihood of choosing at random a set which gives a
non-neglectable contribution to the functional integral is zero.

The standard method to tackle this problem is to optimize the Monte Carlo integration
with respect to the combined probability function, i.e. ``unweighting''.
Looking at eq. \ref{Ifunctional} this procedure simply comes down to changing the 
integration measure by redefining the \pdfs \F\ to $\F^\prime$ together with
the detector response function \D\ to $\D^\prime$ such that all the probability 
densities are absorbed in the integration measure. After such a transformation
eqs.~\ref{Ifunctional} and~\ref{Pfunctional} simplify to
\beq
P_{pdf}^{\cal O}(x_e) = 
\left(\int_{V(\F^\prime)}d\,\F^\prime\ P_{exp}^{\cal O}\left(x_e|x_t(\F^\prime)\right)\right)
\int_{V(\D^\prime)} d\,\D^\prime\ .
\eeq
Translating this in the Monte Carlo integration approach now gives an efficiency of one,
i.e. the Monte Carlo sets are ``unweighted'' and eqs.~\ref{Isum} and~\ref{Psum}
transform in the trivial formula
\beq\label{dens}
P_{pdf}^{\cal O}(x_e)\approx\frac{1}{N}\sum_{i=1}^N P_{exp}^{\cal O}\left(x_e|x_t(\F_i^\prime)\right)\ ,
\eeq
where the probability measure is quantified in the density of 
$(\F^\prime_i,\D^\prime_i)$ functional sets.

Note that we could also have optimized with respect to a subset of the experiments,
e.g. experiments 2 through $n$. We would then find
\beq
P_{pdf}^{\cal O}(x_e)\approx\frac{1}{N}\sum_{i=1}^N 
P_{exp\ 1}\left(x_m^{(1)}\Big|\D_i^\prime\left(x_t^{(1)}(\F_i^\prime)\right)\right)
P_{exp}^{\cal O}\left(x_e|x_t(\F_i^\prime)\right)\ .
\eeq
which tells us how to include an additional measurement, not included in the
optimalization procedure. However, care
has to be taken with the efficiency. The effective number of \pdfs used to estimate
the observable is no longer $N$ but $N_{eff}=\epsilon(P_{exp\ 1})\times N$. 
If the efficiency
is too low the Monte Carlo estimate uncertainty on the observable
will become large and dominate over the \pdf uncertainties. 
This can happen if the newly included 
experiment has a superior uncertainty
analysis compared to the previous included experiments in which case one has to increase
the initial number of sets $N$. Or, alternatively, the newly included experiment disagrees
with the previous included experiments in which case the situation cannot be resolved
by increasing $N$. Coercing the experiments together plainly leads to an incorrect result. 

\subsection{Comparison Methodology}

As explained in section 2.3 the predictions are based on a discreet set of \pdfs 
and will build up a probability density in the space of the observable as is expressed in 
eq.~\ref{dens}. Note that the probability function density function of measuring $x_e$
for observable ${\cal O}$, $P^{\cal O}_{pdf}(x_e)$, depends on the experimental
response function of the measurement. Each discreet prediction will be ``smeared''
into a continuous probability function and averaged over all \pdfs in the
set. This means that the particular form of the experimental response function dictates the
minimum number of \pdfs needed for a satisfactory result. A drawback is that each experiment
measuring the observable will have a different $P^{\cal O}_{pdf}$. In more theoretical studies
one often wants a prediction independent of any particular experiment. In that case
we have to substitute for the experimental response function an idealized detector
model. The most straightforward model is the ``perfect'' detector, i.e.
\beq\label{theoryD}
P_{pdf}^{\cal O}(x_e) \approx \frac{1}{N}\sum_{i=1}^N \delta(x_e-x_t(\F_i))
\eeq
where the sum runs over the optimized \pdfs in the set. Such a result leads
to a traditional scatter plot representation of $P^{\cal O}_{pdf}$.
By introducing a averaging parameter $\Delta$ one can get to a more continues result.
One could for instance use a gaussian with a width $\Delta$
for the response function. However, traditionally
one chooses two theta functions to get a histogrammed representation of $P^{\cal O}_{pdf}$.
\beq\label{theoryH}
P_{pdf}^{\cal O}(x_e) \approx \frac{1}{N}\sum_{i=1}^N 
\Theta(x_e-\frac{1}{2}\Delta-x_t(\F_i))\times \Theta(x_t(\F_i)-x_e-\frac{1}{2}\Delta)\ .
\eeq
The drawback of such procedures is that resulting approximation of $P_{pdf}^{\cal O}$ depends on
the averaging procedure.

While the above procedures will give us an approximation for the probability density function
of the observable it does not represent a likelyhood or confidence level. Using $P_{pdf}^{\cal O}(x_e)$
it is pretty straightforward to construct the confidence level. However some care has to be taken
given the potential non-gaussian nature of the experimental response function. 
First we define the parameter independent log-likelyhood
\beq
L^2(x_e)=-2\times\log(P_{pdf}^{\cal O}(x_e)/P_{expmax}^{\cal O}(x_e))\ ,
\eeq
where 
\beq
P_{expmax}^{\cal O}(x_e)=\max_{x_t} P_{exp}^{\cal O}(x_e|x_t)\ ,
\eeq 
is the maximum obtainable probability given an observed value $x_e$. 
For a gaussian response function the log-likelyhood defined here
resorts back to the usual $\chi^2$ definition.
The definition of log likelyhood as a ratio of probability densities
has the property that it is
independent under reparameterizations of the observable.
We now can define the confidence level
\beq\label{CLdef}
CL^{\cal O}(x_e)=\int d\,x\ \Theta(L^2(x)-L^2(x_e))\times P_{pdf}^{\cal O}(x)\ 
\eeq
which is a proper probability. Given the measured value $x_m$, the
confidence level $CL(x_m)$ quantifies the probability that a 
repeat of the experiment will have a worse agreement with the
theory prediction including the \pdf uncertainties.

\section{The Optimized Sets}

The previous section defined the theory prior with the exception of the actual
\pdf parameterization. In subsection 3.1 the parameterization used for this paper
will be discussed. Once that is done we perform in subsection 3.2
the actual optimalization using various combinations of 
proton $F_2$ and \alfa\ measurements.

\subsection{The Parameterization Choice}

Given this is our first venture into the refractory
subject of \pdf determination we choose 
the well established MRST-parameterization~\cite{MRST} as our first guidance. 
This choice implicitly incorporates 
into the theory prior the accumulated knowledge of
many years of \pdf studies which
can be considered a positive feature. 
However, because we have to adhere to the experimental systematic uncertainties and
statistical interpretation of the results this parameterization 
can turn out to be far too restrictive.
Such restrictive parameterizations might 
result in at least an underestimate of the \pdf uncertainties. Or, more seriously,
in discrepancies between the theory and the data 
stemming not from physics but from the parameterization choice. See for
instance ref.~\cite{CTEQparm} where the gluon \pdf parameterization effects on the
one jet inclusive transverse energy distribution are discussed.

For completeness, the explicit MRST-parameterization at the scale 
$Q_0=1$ GeV is given by
\beqn
xu_v &=& A_{u}x^{-\lambda_u} (1-x)^{\eta_u} (1+\epsilon_{u}\sqrt{x}+\gamma_{u}x)\
\nonumber \\
xd_v &=& A_{d}x^{-\lambda_d} (1-x)^{\eta_d} (1+\epsilon_{d}\sqrt{x}+\gamma_{d}x)
\nonumber \\
xg &=& A_{g}x^{-\lambda g} (1-x)^{\eta g}
(1+\epsilon_{g}\sqrt{x}+\gamma_{g}x)\nonumber \\   
xS \equiv 2x (\bar{d} + \bar{u} + \bar{s})
&=&A_{S}x^{-\lambda_S} (1-x)^{\eta_S} (1+\epsilon_{S}\sqrt{x}+\gamma_{S}x)
\nonumber\\
x\Delta \equiv x(\bar{d} - \bar{u}) &=& 
A_{\Delta}x^{-\lambda_\Delta}
(1-x)^{\eta_{S}+2}(1+\gamma_{\Delta}x+\delta_{\Delta}x^{2})\ .
\eeqn
The normalization coefficients $A_u$, $A_d$ and $A_g$ are determined by the
charge and momentum conservation sum rules. Both the charm and bottom quark \pdfs are generated
through perturbative evolution from mass threshold. For the charm quark the threshold
mass is set to 1.5 GeV, while the bottom quark threshold mass is chosen to
be 4.5 GeV. 

To conclude, using the MRST-parameterization the functional \pdf integration of
subsection 2.3 is reduced to an integration over 21 parameters (not counting
the strong coupling constant \alfa\ and squared renormalization/factorization 
scale $\mu$). The prior 
probability distribution of the parameters is chosen to be uniform. Next, we
have to optimize the parameters according to the combined probability 
density function.

\subsection{The $F_2^{proton}$ and \alfa\ Data Optimized Sets}

Given the choice to abstain for the moment from any non-perturbative modeling in the theory
prior, we have to limit ourselves to proton target data. 
The study of heavy nuclei data by introducing a shadowing model is a subsequent step
in the development of the \pdfs and will be taken in another paper. The five deep
inelastic $F_2^P$ experiments we have selected so far are BCDMS~\cite{BCDMS}, 
NMC~\cite{NMC}, H1~\cite{H1}, ZEUS~\cite{ZEUS} and E665~\cite{E665}. For the
\alfa\ measurement we take the value from ref~\cite{alphas}:
$\alfa(M_Z)=0.119\pm 0.003$.  While this value is in fact a world average of sorts, we will
indicate it in the fits as the ``LEP'' experiment. All the experiments are summarized
in table 1 together with the relevant properties. 
\begin{table}\begin{center}\begin{tabular}{|l|c|c|c|r|}\hline
Experiment & Measurement & usable points & used $x$-range & error analysis \\ \hline
BCDMS & $F_2^P$ & 344 & $7.0\times 10^{-2}$-0.75 & gaussian \\
H1 & $F_2^P$ & 188 & $5.0\times 10^{-5}$-0.32 & half gaussian \\
ZEUS & $F_2^P$ & 187 & $6.3\times 10^{-5}$-0.51 & gaussian \\
NMC & $F_2^P$ & 127 & $8.0\times 10^{-3}$-0.50 & gaussian \\
E665 & $F_2^P$ & 53 & $3.7\times 10^{-3}$-0.39& gaussian \\
LEP & $\alpha_S$ & 1 & N.A. & gaussian \\
\hline\end{tabular}
\caption[]{A summary of relevant properties of experiments considered in this paper.}
\end{center}\end{table}

With the exception of H1, 
all other experiments quote a gaussianized analysis of the systematic uncertainties
(i.e. a correlation matrix of the experimental errors is either quoted 
or can be calculated using the published results).
The H1 $F_2^P$ measurement has 5 sources of systematic uncertainties quoted as
$\pm 1\sigma$ deviations where for some of the sources the $\sigma_+$ is different
from the $\sigma_-$. This means that for H1 we use the technique described 
in the previous section and numerically integrate over the systematic uncertainties
by optimizing the \pdf parameters together with the 5 detector model parameters of H1
using as the $P_{sys}(\D)$ detector response function 5 half-gaussians 
depending on the detector model parameters.
For the other 4 experiments the correlation matrix is used, i.e. the
systematic uncertainties are integrated out analytically.
This implies all potential correlations between the different $F_2^P$ 
experiments are ignored. 

Because no higher twist models are in the theory prior we apply a cut on the momentum
transfer and parton fractions in the $F_2^P$ data 
as suggested in ref.~\cite{MRST}:
$Q^2>2$ GeV$^2$ and $W^2=Q^2\times\left(1/x-1\right)>10$ GeV$^2$.

Apart from optimizing the \pdf integration over the 21+1 \pdf parameters
and, for H1, the 5 detector model parameters, we also include the squared 
renormalization/factorization scale as an optimalization parameter. As explained
in subsection 2.1 the reason is twofold. First of all varying the scale amplifies
the \pdf uncertainty stemming from the fixed order perturbative expansion approximation. 
Secondly, the resulting optimized factorization/renormalization scale will indicate
to which degree the theory model is applicable. A very narrow distribution indicating
a failure of the model, a wide distribution indicating the model is appropriate.  
Note that we have chosen the initial squared scale probability distribution
uniform.

To obtain an optimized set we use a Metropolis Algorithm~\cite{metropolis} combined with
a simulated annealing procedure\footnote{The temperature in the annealing
procedure is identical to the tolerance parameter of ref.~\cite{CTEQlagrange,CTEQhessian}.
A value larger than 1 means the experimental uncertainties are amplified.}~\cite{annealing}.
We will discuss some of its potential shortcomings as they can affect 
the results. The first potential problem is the possibility of equivalent disconnected
maxima. While the annealing should guard against finding local maxima and nudge
the Metropolis walker to the global maximum it can happen that more than one maximum
exists of roughly equal probability which are not connected by a likely path. Given
the finite number of steps the Metropolis walker takes this can cause problems as the
generated optimized \pdfs are concentrated on one of the maxima neglecting the others.
A trail-and-error method of choosing different starting points for the Metropolis
walker can be employed. However for a finite number of steps this does not guarantee
the absence of a secondary maximum.
The second problem arises from the choice of step size for the Metropolis
walker. One wants to keep the step size small relative to the width of the probability
function in order to maintain a good efficiency for generating optimized sets. However,
this leads to correlations between \pdfs subsequently generated.
A common sense approach is to pick a small randomly ordered subset 
out of the generated optimized \pdf parameter points. 
As that might be, for complicated topologies,
especially subspace regions of constant probability, a large number of \pdfs can be
required to sample the entire region important to the functional integral.

Note that these potential problems in generating the optimized sets are numerical in nature
and not a shortcoming of the employed method.
Both problems disappear in the limit of an infinite number of steps.
Any finite number of Metropolis steps leaves the
potential for problems as described above.
For now we use the FERMILAB PC-farm~\cite{pcfarm} to generate
optimized sets of 100,000 \pdfs (i.e. \pdf parameters). Out of the 100,000 sets smaller
subsets of 100, 1,000 and 10,000 \pdfs are constructed by random selection from the 
larger list. For most practical applications 100
optimized \pdfs are already enough. For more complicated analysis
additional optimized \pdfs might be necessary.

\begin{table}\begin{center}
\begin{tabular}{|l|c|c|c|c|c|c|}\cline{2-7}
\multicolumn{1}{c|}{}& 
H1 & BCDMS & E665 & ZEUS & NMC & LEP \\ \hline
H1-MRST set & - & 67\% & 21\% & {\red 0.5\%} & {\red $<$0.1\%} & 31\% \\ \hline
BCDMS-MRST set & 85\% & - & 23\% & {\magenta 1.5\%} & {\red $<$0.1\% }& {\red 0.5\%} \\ \hline
E665-MRST set & 30\% & 82\% & - & {\magenta 1.6\%} & {\magenta 1.0\%} & 99\% \\ \hline
ZEUS-MRST set & 22\% & {\red $<$0.1\%} & {\magenta 5.0\%} & - & {\red $<$0.1\%} & 24\% \\ \hline
NMC-MRST set & {\red $<$0.1\%} & 28\% & {\magenta 1.5\%} & 
{\red $<$0.1\%} & - & {\magenta 3.2\%} \\ \hline
\end{tabular}
\caption[]{The confidence level of each experiment given the different sets.
The name of the set is composed of all included experiments and
the \pdf parameterization choice.}
\end{center}\end{table}
The first optimalization to consider are with respect 
to the individual $F_2^P$ experiments. Given
a set optimized to one of the experiments we can calculate the confidence level that
the prediction using this set describe the other four experiments.
The confidence level, as is detailed in section 2.4,
is defined as the probability a repeat of the experiment would 
result in a worse agreement than the current agreement. 
Given this analysis we can conclude
which experiments are compatible and can be considered in combined optimized sets.
The results are listed in table 2. Note that because different 
experiments will have different experimental response functions
and hence different \pdf uncertainties the confidence level of experiment A given
experiment B is different from the confidence level of experiment B given experiment A.
A dramatic point in case can be seen in table 2 between the H1 and ZEUS experiment.
From the results exposed in table 2 we conclude that both ZEUS and NMC cannot be
combined with any other experiment as they produce their own distinct \pdf sets.
Both NMC and BCDMS prefer a much lower value of \alfa\ than the LEP result. 
We can combine H1, BCDMS and E665 as each of these experiments 
predicts the other two with high confidence.

Given the confidence level results we have constructed,
apart from the 5 sets using the individual $F_2^P$ data results, 7 additional
\pdf sets with different combinations of H1, BCDMS, E665 and LEP results.
Note that we combine the \alfa\ measurement with
BCDMS even though the confidence level tells us it is not compatible. While
inconsistent we do this to get \pdf sets with a somewhat more stabilized and
acceptable \alfa. Alternatively, we could have chosen to fix the value of
\alfa\ to a predetermined value in the physics prior.

With the twelve optimized sets we are now ready for comparisons with hadron collider
data. Given this is the first kind of such an endeavor we can expect surprises which
will guide us to the next step in the further development of \pdfs with uncertainties.
In subsequent papers we will explore the phenomelogical implications for many hadron
collider observables.

\subsection{Some \pdf set Results}

\begin{center}\begin{figure}
\includegraphics{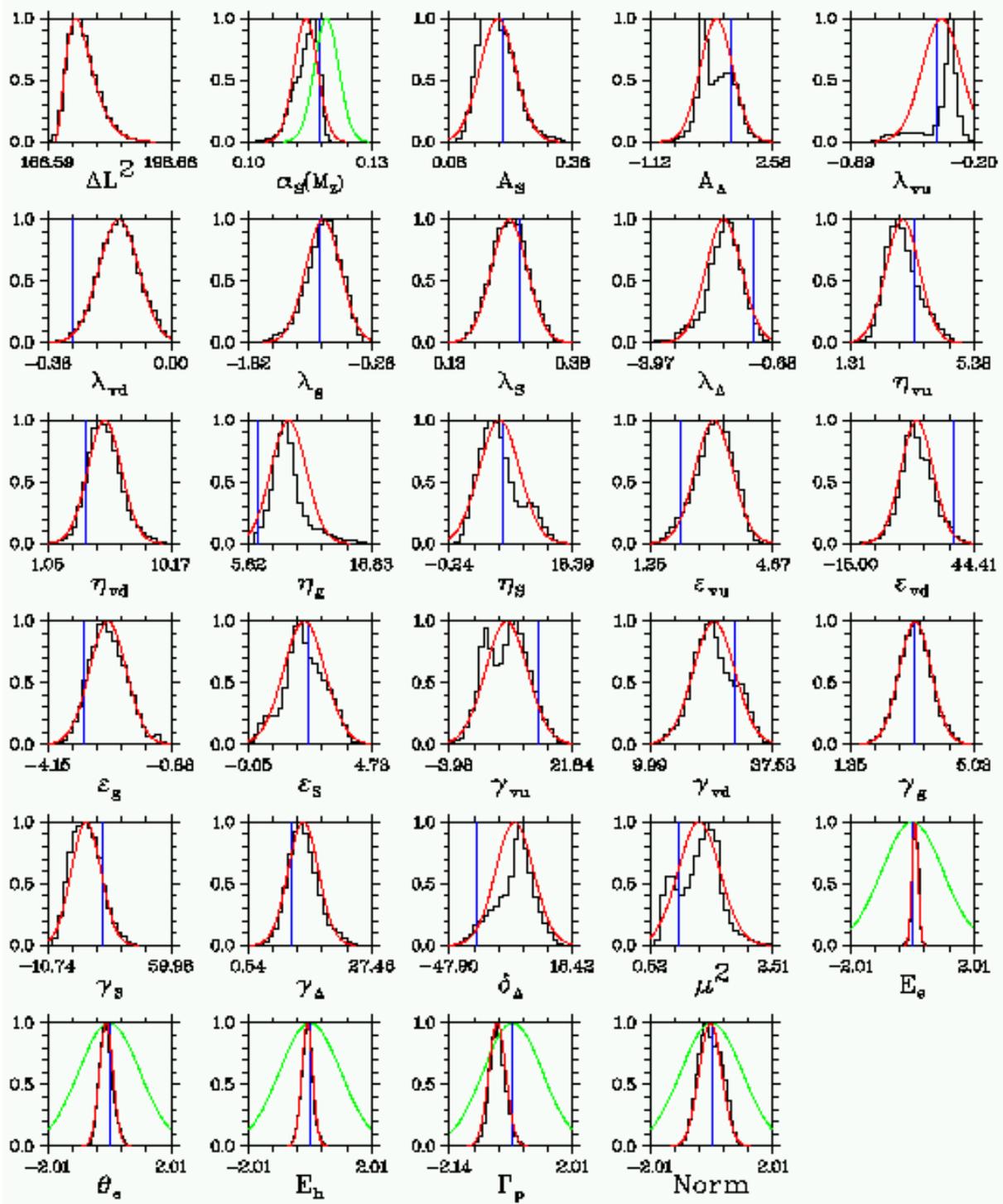}
\vspace{17.0cm}
\caption[]{The H1-MRST set parameter distributions (histograms) together with the
gaussian approximation (red curve) and the central MRST fit result \cite{MRST} (blue bar).
In addition the log-likelyhood $\Delta L^2$ distribution (histogram)
and the $\chi^2$ probability distribution approximation (red curve) is shown. Note that the
last 5 parameters are the detector model parameters of H1 \cite{H1} together
with the prior probability distribution for this parameter (green curve). The green
curve for the \alfa\ result is the LEP value \cite{alphas}}
\end{figure}\end{center}
Given the factorization scheme dependence required for the calculation
of the next-to-leading order matrix elements, 
the individual \pdfs do not contain much physics. Moreover the correlations between
different \pdfs can be expected to be significant given the fact that at leading order
the $F_2^P$ experiments only determine the 
squared electric charge weighted sum of the flavor
\pdfs. At next-to-leading order one can hope at best to distinguish this
charged sum of quark \pdfs from the gluon \pdfb.
This renders a meticulous study of a specific flavor \pdf prodigal.
Yet, a cursory look might be helpful to understand the phenomelogical 
implications for hadron collider observables.

First of all we can look at the 21 \pdf parameters of an optimized set
together with \alfa, the squared renormalization/factorization scale
and the log-likelyhood distribution. In fig.~1 we take as an example the
H1-MRST optimized set which also includes the five detector modeling parameters. 
Prior to the optimalization  each detector modeling parameters is given by a  
gaussian random parameter with an average of zero and a width of one. To obtain
the fractional detector deformation of the $F_2^P$ signal a positive value of the random
variable is multiplied by $\sigma_+$ and a negative value by $\sigma_-$.
Examining fig.~1  we make several general observations which apply also to the other
optimized set results:
\begin{itemize}
\item 
Many of the parameters deviate significantly
from gaussian behavior, letting the linearized error propagation method of 
refs.~\cite{alekhin,botje,zomer} go astray.
\item
The central MRST fit \cite{MRST} is not that far off for each individual 
parameter. The differences are attributed to the different treatment of the
systematic errors and the MRST fit is averaged over many more experiments.
\item 
The optimized detector response parameters are notably close to the H1 central value, 
though some of the distributions are skewed.
\end{itemize}
\begin{table}
\hspace{-1.5cm}\begin{tabular}{|l||c||c|c|c|c|}\cline{2-6}
\multicolumn{1}{c|}{}&\multicolumn{1}{c|}{\alfa-distribution}&
\multicolumn{4}{c|}{$\Delta L^2$-distribution}\\ \hline 
\scriptsize{set} & \scriptsize{CL interval} & 
\scriptsize{$\min(\Delta L^2)$} & \scriptsize{$\langle\Delta L^2\rangle$} & 
\scriptsize{$\frac{1}{2}\sigma^2(\Delta L^2)$} &\scriptsize{NDP/DOF} \\ \hline
\scriptsize{ZEUS-MRST} & \scriptsize{(0.113,(0.114,(0.115),0.116),0.117)} & \scriptsize{466.1} & \scriptsize{481.2} & \scriptsize{42.9} & \scriptsize{187/23} \\ \hline 
\scriptsize{NMC-MRST} & \scriptsize{(0.098,(0.102,(0.108),0.112),0.117)} & \scriptsize{185.4} & \scriptsize{196.4} & \scriptsize{10.1} & \scriptsize{127/23} \\ \hline 
\scriptsize{H1-MRST} & \scriptsize{(0.108,(0.112,(0.115),0.117),0.119)} & \scriptsize{166.6} & \scriptsize{175.9} & \scriptsize{7.8} & \scriptsize{188/28} \\ \hline 
\scriptsize{H1+LEP-MRST} & \scriptsize{(0.114,(0.116,(0.118),0.119),0.121)} &\scriptsize{167.3} & \scriptsize{176.0} & \scriptsize{8.3} & \scriptsize{189/28} \\ \hline 
\scriptsize{BCDMS-MRST} & \scriptsize{(0.104,(0.106,(0.108),0.110),0.112)} &\scriptsize{317.2} & \scriptsize{328.1} & \scriptsize{12.1} & \scriptsize{344/23} \\ \hline 
\scriptsize{BCDMS+LEP-MRST} & \scriptsize{(0.112,(0.113,(0.116),0.117),0.119)} &\scriptsize{325.0} & \scriptsize{335.8} & \scriptsize{15.7} & \scriptsize{345/23} \\ \hline 
\scriptsize{E665-MRST} & \scriptsize{(0.106,(0.112,(0.116),0.127),0.133)} &\scriptsize{57.9} & \scriptsize{65.5} & \scriptsize{4.9} & \scriptsize{53/23} \\ \hline 
\scriptsize{E665+LEP-MRST} & \scriptsize{(0.114,(0.117,(0.120),0.123),0.126)} &\scriptsize{59.1} & \scriptsize{66.5} & \scriptsize{6.0} & \scriptsize{54/23} \\ \hline 
\scriptsize{H1+BCDMS-MRST} & \scriptsize{(0.109,(0.110,(0.112),0.114),0.115)} &\scriptsize{510.9} & \scriptsize{525.8} & \scriptsize{11.4} & \scriptsize{532/28} \\ \hline 
\scriptsize{H1+BCDMS+LEP-MRST} & \scriptsize{(0.110,(0.111,(0.112),0.114),0.115)} &\scriptsize{511.5} & \scriptsize{521.8} & \scriptsize{10.0} & \scriptsize{533/28} \\ \hline 
\scriptsize{H1+BCDMS+E665-MRST} & \scriptsize{(0.109,(0.111,(0.112),0.114),0.115)} &\scriptsize{580.3} & \scriptsize{596.2} & \scriptsize{12.3} & \scriptsize{585/28} \\ \hline 
\scriptsize{H1+BCDMS+E665+LEP-MRST} & \scriptsize{(0.110,(0.112,(0.113),0.114),0.115)} &\scriptsize{579.7} & \scriptsize{592.3} & \scriptsize{10.4} & \scriptsize{586/28} \\
\hline\end{tabular}
\caption[]{The relevant properties of the \alfa\ and $\Delta L^2$ distributions for the
optimized sets. The confidence level intervals are for a CL of 4.55\%, 31.73\% and 100\%.
The bin width used to calculate the confidence level intervals is 0.005 using the 1,000
\pdfs}
\end{table}
Also of interest are parameters not directly related to the \pdfs but important for
the quality of the optimized set. These are the log-likelyhood distribution,
the squared renormalization scale distribution and the \alfa\ distribution for each
of the twelve sets. 

The value of \alfa\ can be used to infer the 
reliability of the experimental error analysis. The 
confidence level intervals of the \alfa\
distribution for all twelve \pdf sets are listed in table 3.
The notation used, $(a,(b,(c),d),e)$, is the following:
the 100\% confidence level interval is given by $c$ (i.e. the maximum value);
the 31.73\% confidence level interval is given by the interval $(b,d)$ (i.e. the ``1-sigma'' interval);
the 4.55\% confidence level interval is given by the interval $(a,e)$ (i.e. the ``2-sigma'' interval).
In order to calculate the confidence level we use eq.~\ref{CLdef} with $P_{pdf}^{\cal O}$
given by eq.~\ref{theoryD}:
\beq\label{CLtheory}
CL(\alfa) = \frac{1}{N}\sum_{i=1}^N\Theta\left(P_{pdf}(\alfa\left(\F_i\right))-P_{pdf}(\alfa)\right)\ .
\eeq
To evaluate the $P_{pdf}$ functions in the step function we have to measure the density of \alfa\ predictions.
To do this we choose the histogram approximation of eq.~\ref{theoryH} with a width of 0.005.
The number of \pdfs involved in the evaluation, $N$, is chosen to be 1,000.
We can see the inclusion of \alfa\ into the H1 and
BCDMS optimalization pulls \alfa\ towards the LEP value. For H1 it more than halves
the uncertainty on \alfa, while for BCDMS the uncertainty is not affected much. This
is understandable because the BCDMS result and the LEP result excluded each other up
to a confidence level of 0.5\%.
Hence, coercing them into a combined optimalization has shifted the optimalization
region and not, as is the case for the H1 optimalization, refined the optimalization 
region. For the sets combining several $F_2^P$ measurements we see that the 
inclusion of the LEP data point hardly affects the already well established
value of \alfa. This can be understood by the large number of datapoints included
in the combined experiments, rendering the uncertainty on \alfa\ significantly
smaller than the uncertainty on the LEP value of \alfa.

From the log-likelyhood we can deduct something about the
quality of the optimalization.
The log-likelyhood distribution is for all twelve sets close to the
expected $\chi^2$ distribution. The relevant parameters are given in table 3.
The minimum log-likelyhood, $\min(\Delta L^2)$, found is on average equal to the number
of data points (NDP) minus the degrees of freedom (DOF). However, due to the
high degree of correlation between the \pdf parameters the {\it effective} degrees
of freedom (i.e. the independent degrees of freedom taking the correlations
into account) is often much lower. The effective degrees of freedom are on average
given by half the squared variance, $\frac{1}{2}\sigma^2(\Delta L^2)$. The difference 
between the average value of the log-likelyhood, $\langle\Delta L^2\rangle$, 
and the minimum value is for a $\chi^2$ distributions equal to
the effective number of independent parameters. From table 3 we can see that
the ZEUS set and to a lesser extend the NMC set are problematic. This does
not come as a surprise given the confidence level results of table 2. 
\begin{center}\begin{figure}[t]
\includegraphics{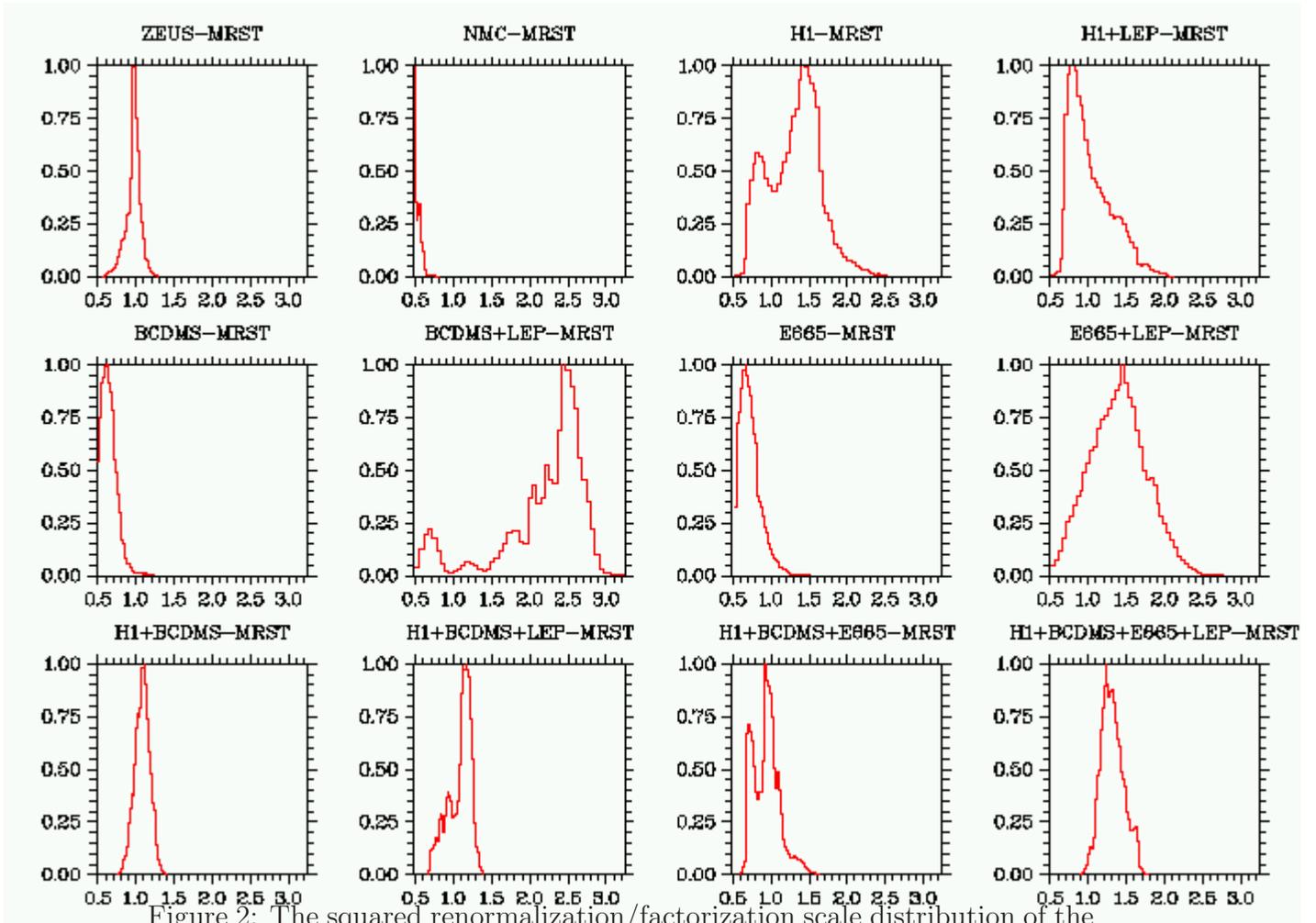}
\vspace{12.0cm}
\caption[]{The squared renormalization/factorization scale distribution of the
optimized sets. The horizontal axis is the squared ratio of the scale and the momentum transfer.}
\end{figure}\end{center}
\begin{figure}[p]
\label{CompPDF}
\begin{center}\vspace{15cm}
\includegraphics{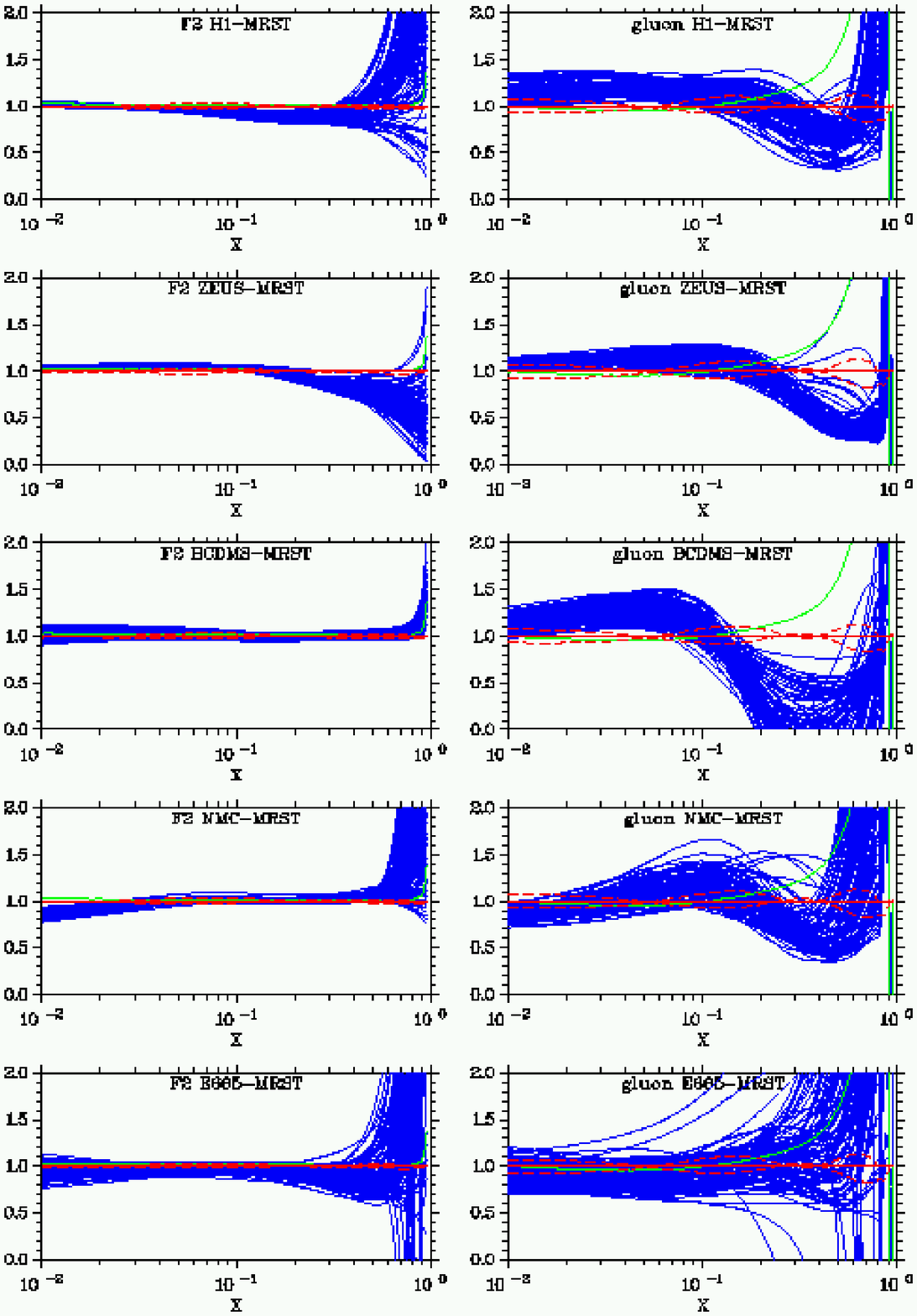}
\end{center}
\caption[]{The ratio of the optimized \pdf sets (blue) and CTEQ5M (green) over MRS99 
at a scale of 10 GeV. The dashed red curves are the two MRS99 sets with \alfa$(M_Z)$ values
of 0.1125 and 0.1225. The left column is the charged summed
combination of \pdfs, while the right column ratio of the gluon \pdf.}
\label{pdfcomp1}
\end{figure}

Using the renormalization/factorization scale dependence 
one can deduce the applicability of the theory model. All 12 distributions are shown
in fig.~2.  The troubled ZEUS and NMC sets are also exposed here by the narrow range
of optimized squared scales signaling a distress between the experimental result and
the theory model. Yet, some of the other distributions are also worrisome indicating
possibly a too restrictive parameterization choice.
Note that while in the prior we assumed an uniform probability distribution for the
squared scale, we restricted the range to be larger than half the squared momentum
transfer $Q^2$ as we have a kinematic cut of $Q^2>2$ GeV$^2$ and a parameterization scale
of $Q_0^2=1$ GeV$^2$.

Finally in fig.~3 we compare the optimized sets with the MRS99~\cite{mrs99} 
and CTEQ5M~\cite{cteq5m} distributions. This
comparison forecasts the deviations we will see in the hadron collider observables
between the global fitter predictions and the optimized \pdf sets. As can be seen
the squared charged sum of the flavored \pdfs agrees pretty well with MRS99 and CTEQ5M up to
the point where there is data. Some subtle differences are present in some of the sets.
The big difference however is the gluon distribution which is much lower above a 
parton fraction of 0.1 and higher between 0.01 and 0.1.
These differences are consistent
for all optimized \pdf sets using the $F_2^P$ data and 
will be reflected in the phenomenology at hadron colliders.

It is interesting to note that the differences in the gluon \pdf are especially large
when compared with CTEQ5M. This gives us a first hint at the origin of differences
with the optimized sets. The relevant difference between MRS99 and CTEQ5M 
is the choice of experiment to constrain the gluon \pdf in the large $x$ region.
The MRS99 global fit makes the traditional choice of using the WA70 \cite{WA70} prompt photon data. 
Because the prompt photon data suffers from large theoretical uncertainties the CTEQ5M
global fit instead chose the one jet inclusive transverse jet energy data of CDF
and D0 \cite{CDFonejet,D0onejet}. Yet the deep inelastic proton data favors the
MRS99 over the CTEQ5M global fit result for large $x$ gluons. This is a rather
troubling conclusion as it suggests a very significant deviation between 
the deep inelastic proton data and high transverse energy jet data at the TEVATRON.
 
\section{Conclusions and Outlook}

In this paper we introduced a flexible method to incorporate \pdf uncertainties
into phenomelogical predictions of collider observables. By representing the
uncertainties as a density, i.e. an ensemble of \pdfs with which one predicts 
the observable, the practical use is both convenient and flexible. The sets used
in this paper can be obtained from the website {\tt pdf.fnal.gov}.

The restriction to a {\it minimum} set of experiments to determine the optimized
\pdfs is deliberate as many experiments do not seem to be consistent with a single
optimized \pdf set. One of the next steps in developing the optimized \pdfs
is to consider deuterium data and the related nuclear effects in the optimalization. 
While for deuterium the nuclear effects are small and manageable within the context
of \pdf optimalization, one can expect serious complications if one considers
heavier targets. It might be better to use the optimized \pdfs to {\it predict} these
measurements in order to obtain better understanding of the nuclear effects.

In the current optimized sets the most severe prior assumption is due to the
restrictive MRST parameterization. One can argue the seemingly discrepancies
between proton $F_2$ measurements and high energy jet production at CDF and
D0 is due to the restrictive parameterization. It is therefore crucial to device
more general schemes for parameterizing the \pdf functionals. One way is to use
complete sets of functions, in such a way 
that the higher order functions are associated with
smaller scale fluctuation. The drawback of such methods is that changes in a
parameter affect the \pdf for all values of the parton fractions $x$. This induces
strong correlations between the parameters and more importantly accurate data
in localized regions of $x$ will affect the uncertainty estimates for all regions of 
$x$. A better method would be splines or other methods of generating functions using
the value of the \pdf at fixed grid points $x_i$. In such a case the parameters
(i.e. the value at the grid points) are more localized. Such methods are under
development and the subject of future publications.

Using the current optimized \pdf sets a series of studies on hadron collider observables
will be published. These studies range from luminosity measurements using \W and \Z cross
sections to jet production. Given the upcoming run II at Fermilab and the LHC program at
Cern further development of the optimized \pdfs is more than warranted as it will have
a large impact on the phenomenology at these hadron colliders.

\end{document}